# One Pot Synthesis of Cubic Gauche Polymeric Nitrogen

Runteng Chen[1,2] †, Jun Zhang[1] †*, Zelong Wang[1], Ke Lu[1], Yi Peng[1], Jianfa Zhao[1], Shaomin Feng[1,2] & Changqing Jin*[1,2]

[1]Beijing National Laboratory for Condensed Matter Physics, Institute of Physics, Chinese Academy of Sciences, Beijing 100190, P. R. China.

[2]School of Physics, University of Chinese Academy of Sciences, Beijing 100190, P. R. China.

†These authors contributed equally.
*Corresponding author: zhang@iphy.ac.cn; jin@iphy.ac.cn

**Abstract**

The long sought cubic gauche polymeric nitrogen (cg-N) consisting of N-N single bonds has been synthesized by a simple route using sodium azide as a precursor at ambient conditions. The recrystallization process was designed to expose crystal faces with low activation energy that facilitates initiating the polymeric reaction at ambient conditions. The azide was considered as a precursor due to the low energy barrier in transforming double bonded N=N to single bonded cg-N. Raman spectrum measurements detected the emerging vibron peaks at 635 $cm^{-1}$ for the polymerized sodium azide samples, demonstrating the formation of cg-N with N-N single bonds. Different from traditional high pressure technique and recently developed plasma enhanced chemical vapor deposition method, the route achieves the quantitative synthesis of cg-N at ambient conditions. The simple method to synthesize cg-N offers potential for further scale up production as well as practical applications of polymeric nitrogen based materials as high energy density materials.

**Introduction**

The N≡N triple bond is one of the strongest chemical bonds, with a very high bond energy of 954 kJ/mol [1], while the N-N single bond is much weaker, with a bond energy of 160 kJ/mol [2, 3]. Due to the huge difference in energy between triply bonded dinitrogen and singly bonded nitrogen, the polymerized nitrogen materials with N-N single bonds are proposed as high energy density materials and expected to release a high amount of chemical energy while transforming into a triple bonded dinitrogen molecules. It was anticipated that the greatest utility of fully single-bonded nitrogen would produce a tenfold improvement in detonation pressure over HMX [4] and therefore polymerized nitrogen is highly sought as a high energy density material.

It was predicted in 1985 that molecular nitrogen would polymerize into atomic solid at high pressure [5]. Later, the cubic gauche (cg-N) lattice with a single bonded diamond like structure was proposed [6]. Subsequently, large amounts of theoretical work reported on the high pressure approach to polymerized nitrogen with N-N single bonded forms [7-23]. It was not until 2004 that Eremets *et. al.* prepared cg-N directly from molecular nitrogen at pressures above 110 GPa (1GPa≈10,000 atmospheric pressure) and temperatures above 2000 K using laser heated diamond cell technique [24]. The cubic gauche structure was confirmed by growing the single crystal in 2007 [25]. However synthesis pressures above 110 GPa are highly required. Later on polymerized nitrogen with layered structure (LP-N) [26], hexagonal layered structure (HLP-N) [27], black phosphorus nitrogen (BP-N) [28], and Panda nitrogen [29] were successfully prepared at 120 GPa, 244 GPa, 146 GPa & 161 GPa, respectively. Nevertheless, the high pressure is commonly regarded as a necessary route to synthesize polymerized nitrogen. All these polymerized nitrogen samples decomposed in the process of releasing pressure. Some strategies, such as chemical doping, were attempted at high pressure by introducing metals in order to stabilize the polymerized nitrogen lattice at lower pressure [30-34] but it remains challenging to recover the polymerized nitrogen samples at ambient pressure. In addition, introducing non-energetic element, especially the heavy metal seriously decreases energetic density of polymerized nitrogen materials, detrimental to the natural applications.

In the past few years, plasma enhanced chemical vapor deposition (PECVD) technique has been developed to take advantage of the non-equilibrium plasma environment to synthesize polymerized nitrogen materials. The relative thermodynamic instability of polymerized nitrogen can be overcome by kinetic stability provided through energy band hybridization by a charge transfer mechanism, which makes the high energy lattice gain stability and avoid the highly energetic conversion to molecular $N_2$ [35]. The cg-N and $N_8$ cluster were successfully prepared with or without the assistance of carbon nanotube using sodium azide as precursor [35-41]. However the plasma generally can only penetrate several nanometers into the solid surface [42], leading to limited quantity of cg-N. Besides, high energy plasma particles also could dissociate the cg-N produced in the experiment, which is backed up by the fact that the optimized deposition time is demanded for a very small amount of cg-N sample prepared by the PECVD method [38].

Theory proposes that the transition path to polymerized nitrogen might pass through different molecular structures with limited thermodynamic stability fields[43]. Additionally earlier experiments aimed at synthesizing polymerized nitrogen only reached the amorphous state at ambient temperature compression, indicating that increasing reaction temperature to activate the precursor might play a crucial role in the formation of the atomic structures [44, 45]. For example $LiN_5$ and $K_2N_6$ were synthesized at lower pressures [30, 32, 33] with the assistance of laser heating.

We here for the first time report a simple route to synthesize cg-N by using sodium azide as a precursor. The cg-N was successfully obtained through recrystallization reaction, followed by transforming azide ions with N=N double bonds into single bonded cg-N that can stabilize at ambient pressure conditions. To date it is the simplest synthesis route to realize cg-N that can be further developed for scale up production and potential practical applications.

**Materials & Methods**

High pure sodium azide ($NaN_3$) (≥99.9%) was chosen as nitrogen source to synthesize cg-N. The recrystallization process was performed first. A solution of $NaN_3$ (2 mol/L) was pre-prepared and dropped into a crucible. Then the crucible was transferred into a tubular furnace. The vacuum operation was performed to realize the recrystallization of the $NaN_3$ powder. The reaction temperatures and times within the 200-300°C and up to 10 hours were tried in the synthesis processes. The optimized synthesis condition with reaction temperature 240 ~ 260 °C and heating time 5 hours was employed to synthesize the cg-N. The sample was named as polymerized sodium azide followed by synthesis temperature, time, such as PSA-240°C-5h. According to aforementioned conditions, heating $NaN_3$ without recrystallization process to prepare cg-N was tried and it was named as unpolymerized sodium azide (UPSA).

The Raman spectra were recorded in range of 100~1500 $cm^{-1}$ with a spectral resolution of 1 cm$^{-1}$ using an integrated laser Raman system (Renishaw) with a confocal microscope and multichannel air cooled CCD detector. The excitation source is an Ar ion laser ($\lambda_0$ =532 nm).

**Results**

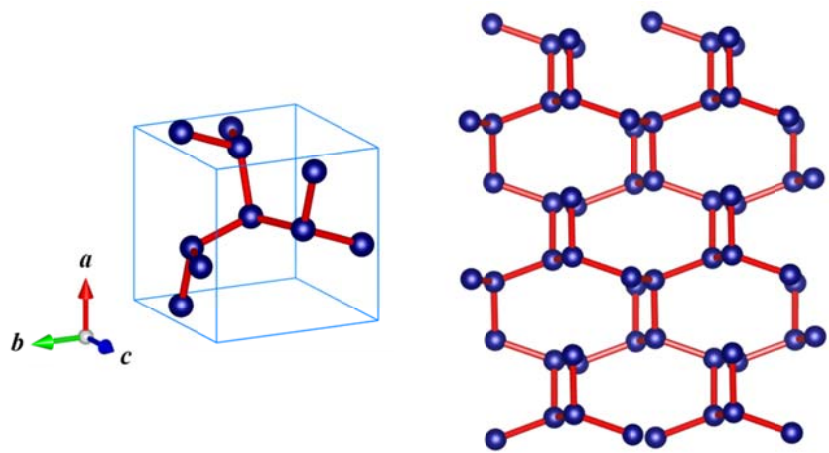

**Fig. 1 The schematic view of cg-N crystal structure.**

The cg-N was proposed to adopt a body centered cubic Bravais lattice, belonging to $I2_13$ space group with a lattice constant of 3.773 Å. The N-N bond length is predicted to be 1.40 Å and the bond angle is 114.0° at ambient pressure conditions [6, 46]. Fig. 1 shows the crystal structure of cg-N, in which each nitrogen atom bonds to three nearest neighbors with N-N single bonds at equal distances and the remaining *p* orbital is occupied by the lone pair electrons. Thus cg-N exhibits a near tetrahedral structure similar to that in diamond (left in Fig.1). The single bonded nitrogen atoms form fused rings, which connect with others forming a three dimensional network structure [47], as shown in the right panel of Fig. 1. The theoretical work on cg-N didn't show imaginary frequency in phonon spectrum calculation [12], indicating that the entire covalent network of cg-structure is metastable at ambient pressure.

The azide ion is composed of N=N double bonds that can be used as an ideal precursor for synthesizing the cg-N due to the low energy barriers for transforming into single bonded cg-N. In this work the sodium azide is used in present simple experiments. The recrystallization process is expected to exposing some crystal faces with low activation energy. Then the polymeric reaction can be initiated at ambient conditions. To date the amount of cg-N obtained experimentally at ambient pressure is still not sufficient to obtain ideal X-ray diffraction patterns. Therefore the Raman spectrum was adopted to characterize the chemical bond transformation in the present work that is commonly used for experimental detections of polymerized nitrogen.

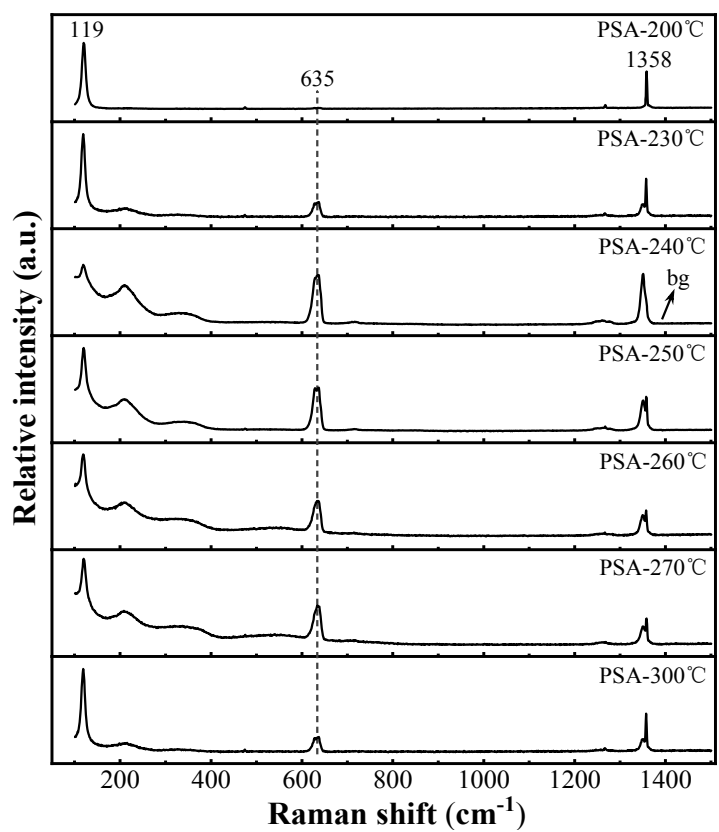

**Fig. 2 The Raman spectra of polymerized sodium azide (PSA) synthesized at 200 ~300 °C with a reaction time of 3 hours, where the fingerprint peak at 635 cm$^{-1}$ for cg-N is detected. The vibron intensity of cg-N shows optimized synthesis temperature between 240~260 °C.**

Fig. 2 shows the Raman spectra of polymerized NaN$_3$ (PSA) synthesized at 200~300 °C with a reaction time of 3 hours. The sharp peaks of NaN$_3$ near 120, 1267 & 1358 cm$^{-1}$ are assigned to the vibrational lattice mode, the first overtone of the IR active bending ν2 mode and the symmetric stretching ν1 mode of the azide ion, respectively, which matches well with the Raman spectra of free standing NaN$_3$ [41]. For comparison the NaN$_3$ raw material and a sample heated only at 250 °C for 3 hours without the recrystallization process were characterized and showed same Raman vibron profiles without any signals at 635 cm$^{-1}$ (see Fig. S1). The lines at 635 cm$^{-1}$ for polymerized sodium nitrogen sample (PSA) are comparable to the theoretical results extrapolated to ambient pressure because the Raman modes tend to soften with decreasing pressure [24, 46, 48]. It also agrees with our recent work on polymerized cg-N based on potassium azide. Therefore the emergent intense vibron at 635 cm$^{-1}$ is assigned to the pore breathing A symmetry and is regarded as the finger print peak of cg-N, which unambiguously indicates the successful synthesis of cg-N [46].

The synthesis conditions of cg-N were first systematically investigated by controlling the reaction temperature with a fixed reaction time of 3 hours. As shown in Fig. 2, the vibron intensity of cg-N increases as the reaction temperature raises from 200 $^{o}$C to 240 $^{o}$C compared to that at 1358 cm$^{-1}$ for the unreacted or remnant NaN$_3$ in the PSA samples. As temperature increases up to 270 $^{o}$C, the Raman intensity of cg-N remains comprehensive but decreases clearly at 300$^{o}$C. Additionally, it is worth noting that the intensity of vibrational lattice mode at 120 cm$^{-1}$ for the unreacted NaN$_3$ in PSA-240 sample dramatically decreases compared to those prepared at other temperatures. The PSA sample exhibits uniformly dark blue color and should be related to the by product formed during the synthesis of cg-N and will be discussed below. Therefore, the optimized temperature range of 240~260 $^{o}$C was adopted for cg-N preparation.

A quantitative comparison was made by calculating the intensity ratios of the characteristic Raman peaks to measure the abundance of cg-N phase in the PSA samples. Table I summarizes the partial line intensities of NaN$_3$ and cg-N. The amount of cg-N synthesized is reflected by the peak intensity at 635 cm$^{-1}$, while those at 120 and 1358 cm$^{-1}$ represent the unreacted NaN$_3$. Herein, a simplified conversion degree from NaN$_3$ to cg-N is defined by the peak intensity ratio. The intensities of NaN$_3$ at 120 cm$^{-1}$ and 1358 cm$^{-1}$, calculated by $(I_{200}-I_{bg})/(I_{1358}-I_{bg})$ for the NaN$_3$ raw material (SA-R), NaN$_3$ heated only (UPSA) and polymerized NaN$_3$ (PSA) samples at different temperatures are between 2 and 3 and remain little changed. For the SA-R and UPSA samples no trace of cg-N is detected. After recrystallization process the vibron intensity of cg-N enhances dramatically for PSA samples and the ratio of cg-N to NaN$_3$ is 1.26-1.39 calculated by $(I_{635}-I_{bg})/(I_{1358}-I_{bg})$, and 0.42~ 0.60 calculated by $(I_{635}-I_{bg})/(I_{120}-I_{bg})$. Therefore it indicates that the crystallization process strongly promotes the formation of cg-N and the optimized synthesis temperature range is 240~260 $^{o}$C.

**Table I The main Raman peak intensity for sodium azide (NaN$_3$) raw material (SA-R), NaN$_3$ heated only (UPSA) and polymerized NaN$_3$ (PSA) samples at 200 ~ 300 $^{o}$C for 3 hours.**

| Sample | $I_{bg}$ | $I_{120}$ | $I_{635}$ | $I_{1358}$ | $(I_{120}-I_{bg})/(I_{1358}-I_{bg})$ | $(I_{635}-I_{bg})/(I_{120}-I_{bg})$ | $(I_{635}-I_{bg})/(I_{1358}-I_{bg})$ |
|---|---|---|---|---|---|---|---|
| SA-R | 1401 | 61101 | / | 23411 | 2.71 | | / |
| UPSA-250℃ | 505 | 18324 | / | 8093 | 2.35 | | / |
| PSA-200℃ | 189 | 3019 | | 1192 | 2.82 | | |
| PSA-230℃ | 225 | 4800 | 964 | 2298 | 2.21 | 0.16 | 0.36 |

| Sample | | | | | | | |
|---|---|---|---|---|---|---|---|
| PSA-240℃ | 503 | 8325 | 5185 | 4231 | 2.10 | 0.60 | 1.26 |
| PSA-250℃ | 3382 | 48310 | 26556 | 21578 | 2.47 | 0.52 | 1.27 |
| PSA-260℃ | 1100 | 12360 | 5834 | 4509 | 3.30 | 0.42 | 1.39 |
| PSA-270℃ | 4952 | 46283 | 17281 | 15951 | 3.76 | 0.30 | 1.12 |
| PSA-300℃ | 1996 | 49297 | 3205 | 20430 | 2.57 | 0.03 | 0.07 |

Notes: Table I shows the intensities ($I$) of Raman vibrons at 120, 635 and 1358 $cm^{-1}$. Background is defined by the intensity of linear part at 1400 $cm^{-1}$ marked in the Fig. 2.

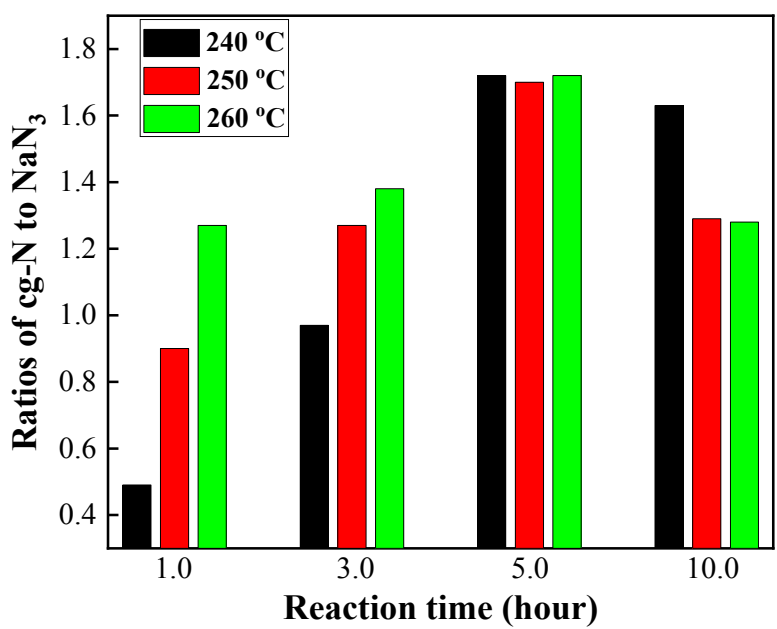

**Fig. 3 The reaction time dependence of the conversion ratios of cg-N to $NaN_3$ for the PSA samples prepared at 240, 250 & 260 $^oC$. The optimized synthesis conditions for the PSA samples are 240~260 $^oC$ with a reaction time of 5 hours.**

Further the preparation conditions were explored by synthesizing PSA samples at 240~260 $^oC$ with variable reaction time. The Raman vibron intensity of cg-N in the PSA-240 samples increases gradually as the reaction time extends up to 5 hours and then decreases, which is shown in Fig. S2. Samples prepared at 250 and 260 $^oC$ show the same results (Fig. S3). For comparison, the intensity ratios of the cg-N to unreacted $NaN_3$ in theses PSA samples were calculated in Table SI. The synthesis time dependence of the ratios of cg-N to $NaN_3$ at 240-260 $^oC$ was plotted in Fig. 3. All samples show consistent trends: the PSA samples with a reaction time of 5 hours show the highest conversion degree. Therefore, the optimized preparation conditions for polymerized sodium azide samples are a temperature range of 240~260 $^oC$ and a reaction time of 5 hours.

**Discussion**

To date for cg-N preparation the high pressure technique and PECVD method have been proposed, which suffer from the challenge that samples could not recover to ambient pressure and enhancing yield is difficult. Here a simple route was designed for synthesizing cg-N under atmospheric conditions considering the thermodynamic barriers related to transforming to N-N single bonds and activation energy of precursor. In our experiments it was found that recrystallization process played a crucial role to form cg-N. For azides the crystal face with low activation energy may facilitate initiating the transformation of the double bonds N=N in azide ions to single bonded cg-N. The optimized temperatures for cg-N preparation by $NaN_3$ are primarily restricted by the melting points $NaN_3$ (~275°C). These results align well with the fact that the cg-N forms on the crystal face with lower activation energy of azides precursors oriented by the recrystallization process. Regarding the formation mechanism of cg-N it was observed that the polymerized sodium azide turned dark blue in our experiments that should be related to the formation of by-product $Na_3N$ during the synthesis of cg-N. This is consistent with the reported results that pH value of the solution increases after synthesis reaction as $Na_3N$ displays alkalinity when dissolved in water [38]. A more detailed conversion mechanism is currently at investigation.

**Conclusion**

The polymeric nitrogen with cubic gauche structure was successfully synthesized by a simple route using sodium azide as a precursor. A high yield may be obtained by separating the by products formed on the surface of the produced cg-N. The route is promising for further scale up or applied to synthesize other metastable materials.

**Conflict of Interest**

There are no financial conflicts of interest to disclose.

**Acknowledgments**

The work was supported by the National Natural Science Foundation of China, Ministry of Science & Technology, and CAS Project for Young Scientists in Basic Research of China.

**Author Contributions**